\documentclass[reprint, superscriptaddress, secnumarabic, amssymb, nobibnotes, aps, prl]{revtex4-2}

\pdfoutput=1

\usepackage{tikz}
\usetikzlibrary{positioning}
\usepackage{graphicx}
\usepackage{epstopdf}
\usepackage[T1]{fontenc}
\usepackage[utf8]{inputenc}
\DeclareUnicodeCharacter{2212}{-}
\DeclareUnicodeCharacter{0327}{-}
\usepackage[utf8]{inputenc}
\usepackage{amsbsy}
\usepackage{gensymb}
\usepackage[T1]{fontenc}
\usepackage{amsmath}
\usepackage{amssymb}
\usepackage{bbm}
\usepackage{physics}
\usepackage{multirow}
\usepackage{braket}
\usepackage{xcolor}
\usepackage{comment}
\allowdisplaybreaks
\usepackage{graphicx}
\usepackage[colorlinks=true]{hyperref}  
\hypersetup{
    bookmarks=true,         % show bookmarks bar?
    unicode=false,          % non-Latin characters 
    pdftoolbar=true,        % show Acrobat
    pdfmenubar=true,        % show Acrobat 
    pdffitwindow=false,     % window fit to page when opened
    pdfstartview={FitH},    % fits the width of the page to the window
    pdftitle={BiPt Superconducting properties},    % title
    pdfauthor={S. Sharma}, %aut
    pdfsubject={},   % subject of the document
    pdfcreator={},   % creator of the document
    pdfproducer={}, % producer of the document
    pdfkeywords={BiPt}, % list of keywords
    pdfnewwindow=true,      % links in new window
    colorlinks=true,       % false: boxed links; true: colored links
    linkcolor=blue, %red,          % color of internal links (change box color with linkbordercolor)
    citecolor=blue,        % color of links to bibliography
    filecolor=magenta,      % color of file links
    urlcolor=blue}            % color of external links
\usepackage[normalem]{ulem}
\usepackage{orcidlink}
\renewcommand{\approx}{\simeq}

\begin{document}
    
\preprint{APS/123-QED}

\title{Conventional superconductivity in single-crystalline BiPt  }% Force line breaks with \\

\author{S.~Sharma\,\orcidlink{0000-0002-4710-9615}}
\affiliation{Department of Physics and Astronomy, McMaster University, Hamilton, Ontario L8S 4M1, Canada}
\email[]{shar0486@umn.edu}
\affiliation{
School of Physics and Astronomy, University of Minnesota, Minneapolis, MN 55455, USA}
\author{M.~Pula\,\orcidlink{0000-0002-4567-5402}}
\affiliation{Department of Physics and Astronomy, McMaster University, Hamilton, Ontario L8S 4M1, Canada}
\author{Sajilesh K. P.\,\orcidlink{0000-0002-9921-4062}}
\affiliation{Physics Department, Technion-Israel Institute of Technology, Haifa 32000, Israel}
\author{J. Gautreau}
\affiliation{Department of Physics and Astronomy, McMaster University, Hamilton, Ontario L8S 4M1, Canada}
\author{B. S. Agboola}
\affiliation{Department of Materials Science and Engineering, McMaster University, 1280 Main St W, Hamilton, L8S 4L8, ON, Canada}
\author{J. P. Clancy}
\affiliation{Department of Physics and Astronomy, McMaster University, Hamilton, Ontario L8S 4M1, Canada}
\author{J. E. Sonier}
\affiliation{Department of Physics, Simon Fraser University, Burnaby, British Columbia, Canada, V5A 1S6}
\author{A. Ghara}
\affiliation{Indian Institute of Science Education and Research Pune, Pune, 411008, India}
\author{S. R. Dunsiger}
\affiliation{TRIUMF, Vancouver, British Columbia V6T 2A3, Canada}
\author{M. Greven}
\affiliation{
School of Physics and Astronomy, University of Minnesota, Minneapolis, MN 55455, USA}
\author{M. J. Lagos}
\affiliation{Department of Materials Science and Engineering, McMaster University, 1280 Main St W, Hamilton, L8S 4L8, ON, Canada}
\author{
A. Kanigel}
\affiliation{Physics Department, Technion-Israel Institute of Technology, Haifa 32000, Israel}
%\author{E.~S. S\o rensen\,\orcidlink{0000-0002-5956-1190}}
%\affiliation{Department of Physics and Astronomy, McMaster University, Hamilton, Ontario L8S 4M1, Canada}
\author{G.~M.~Luke\,\orcidlink{0000-0003-4762-1173}}

\email[]{luke@mcmaster.ca}
\affiliation{Department of Physics and Astronomy, McMaster University, Hamilton, Ontario L8S 4M1, Canada}
\affiliation{TRIUMF, Vancouver, British Columbia V6T 2A3, Canada}

\date{\today}% It is always \today, today,

\begin{abstract}
\begin{flushleft}
\end{flushleft}

Binary Bi-Pd/Pt systems have attracted a lot of interest because of their topologically non-trivial nature along with superconductivity. We report the structural and superconducting properties of high-quality single-crystalline BiPt using a comprehensive range of experimental techniques, including X-ray diffraction, electron microscopy, muon spin rotation/relaxation ($\mu$SR), magnetization, resistivity, and heat capacity. Our findings establish that BiPt is a weak type-II superconductor with a transition temperature (T$_c$) of 1.2 K which exhibits pronounced anisotropic superconducting characteristics attributed to its hexagonal crystal structure. Magnetization and electronic transport studies reveal that BiPt lies within the dirty limit, while $\mu$SR and heat capacity data indicate conventional $s$-wave superconductivity that maintains time-reversal symmetry. This work provides valuable insights into the pairing symmetry and superconducting mechanism of topologically trivial BiPt, a sound comparison system for other Bi-based topologically nontrivial superconductors. 

\end{abstract}
\maketitle

\section{\label{sec:level1}INTRODUCTION \protect\\ }
Materials with nontrivial band structures are being studied intensively because of their ability to host novel states like topological superconductivity. Topological superconductors (TSCs), widely known for their potential application in quantum computation \cite{lian2018topological}, are characterized by a bulk superconducting gap with Majorana zero modes in the topological surface states protected by symmetry ~\cite{Fu_Kane_2008}.  Binary Bi-based superconductors have garnered attention for their potential to realize topological superconductivity~\cite{sun2015dirac, Benia_Wahl_2016, Sakano2015, Lv_Xue_2017}. These materials, which have relatively large spin-orbit coupling, can host continuous, direct band gaps and band inversions, allowing for nontrivial band topologies. Dirac surface states observed in $\alpha$-BiPd \cite{sun2015dirac} and  $\alpha$-Bi$_2$Pd \cite{Benia_Wahl_2016}, and triple-degenerate point fermions in t-PtBi$_2$ \cite{gao2018}, and $\beta$-Bi$_2$Pd \cite{Sakano2015} are a few examples of the Bi-based materials with nontrivial band topologies. 

Bi-based Pd/Pt systems display a rich diversity in structural, superconducting, and topological properties. For instance, t-PtBi$_2$ (hexagonal) exhibits surface superconductivity with a transition temperature above 5 K \cite{schimmel2024}, while $\beta$-Bi$_2$Pt (cubic) is non-superconducting. Both phases show large unsaturated magnetoresistances \cite{wu2020GMR, Gao2017GMR}. $\alpha$-Bi$_2$Pd (monoclinic) hosts Dirac surface states alongside $s$-wave superconductivity \cite{dimitri2018}, whereas $\beta$-Bi$_2$Pd (tetragonal) exhibits topological surface states and a triplet component in superconductivity \cite{li2019, Sakano2015}. Similarly, $\alpha$-BiPd (monoclinic) hosts spin-polarized topological surface states \cite{neupane2016}, whereas  $\gamma$-BiPd (hexagonal) displays singlet superconductivity and Dirac surface states\cite{chiang2023,sharma2024}. 
 
BiPt is isostructural to $\gamma$-BiPd, which was predicted to host topological superconductivity due to its non-trivial band structure and $s$-wave superconductivity \cite{sharma2024}. Although, the superconductivity in BiPt was previously identified \cite{Matthias1953, zhuravlev1959}, the nature of the superconducting ground state was not studied in detail. Here, we report the structural, electronic, and thermodynamic properties of single-crystalline BiPt in superconducting and normal states using bulk and microscopic probes such as muon spin relaxation and rotation ($\mu$SR).  We observe that the BiPt hosts time reversal symmetry preserving bulk superconductivity of weak type-II nature, where the superconducting gap function is $s$-wave type.

%%-------------------FigureBegin-----------------------%%
\begin{figure}[]
\begin{center}
\includegraphics[scale=0.4]{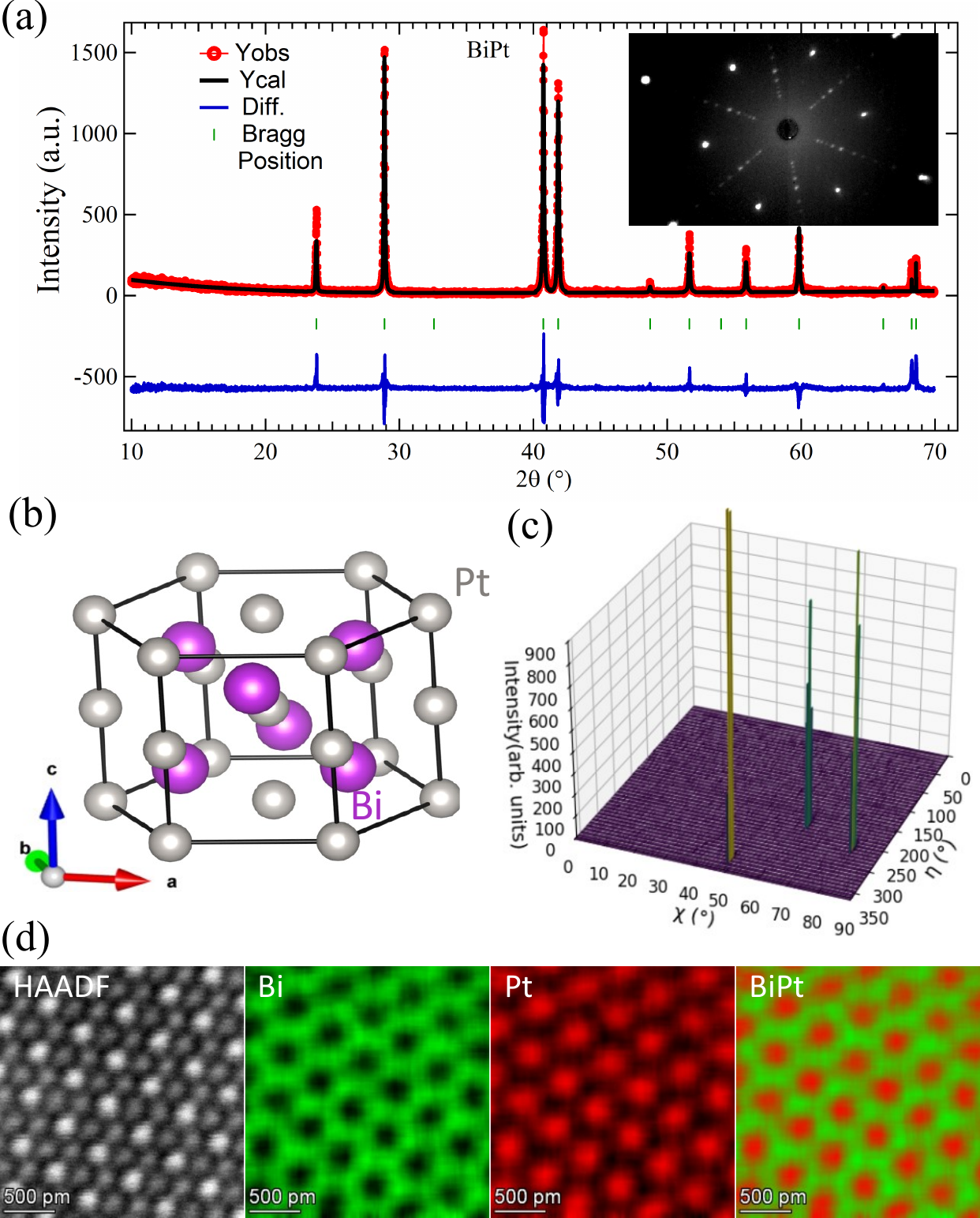}
\caption{Structural characterization of BiPt. (a) Powder XRD pattern confirms that BiPt crystallizes in a hexagonal crystal structure with space group P6$_3$/mmc (194) [Inset] Laue back reflection spectra of <001> plane clearly shows a six-fold symmetry as expected, confirming the high quality of the crystal. (b) Crystal structure of BiPt with Pt forming the hexagonal plane.   (c) Plot of the neutron pole figure measurement performed on a 5 g piece of BiPt is shown. The data correspond to the (1,0,0)/(0,1,0) Bragg peak, which has a multiplicity of six.  The three strong peaks shown here are consistent with one dominant grain, confirming the single crystallinity of the modified Bridgeman-grown sample. (d) Comparison of HAADF STEM with the EDX image which confirms single crystallinity and phase purity at the microscopic level. The HAADF STEM image of the <001> plane of BiPt shows a sharp contrast between Bi and Pt elements forming the hexagons. \label{fig1} }

\end{center}
\end{figure}
%%------------------FigureEnd-----------------------%%
%%%%%%%%%%%%%%%%%%%%%%%%%%%%%%%%%%%%%%%%%%%%%%%%%%%%%%%%%%%%%%%%%%%%%%%%%%%%%%%%%%%%
\section{Experimental Details}
The single-crystal samples of BiPt were grown in a floating-zone furnace using the Bridgman method. High-purity pieces of Bi and Pt were sealed in a quartz tube with a conical bottom. This was first heated to 1000$\degree$C in a box furnace (to make a homogeneous mixture), then melted ($T$>$T_{melt}$)  in a floating zone furnace and slowly lowered through the hot zone at 0.5 mm/h, resulting in a 2 cm long conical single-crystal weighing more than 5 g. 

We confirmed phase purity by X-ray diffraction and crystallinity by neutron pole figures and Laue back diffraction, as shown in Fig: \ref{fig1}.   We crushed a small single crystal piece into powder and measured the powder x-ray diffraction using a PANalytical XPert$^3$ powder diffractometer. We measured neutron pole figures on the 2 cm long crystal via the McMaster Alignment Diffractometer (MAD) located on Beamport 6 in the McMaster Nuclear Reactor.  

High angle annular dark field (HAADF) scanning transmission electron microscopy (STEM) imaging and energy dispersive X-ray spectroscopy (EDS), as shown in Fig. \ref{fig1}(d), were performed using a Spectra Ultra STEM operated at 200 kV. The electron probe had a convergence semi-angle of 28 mrad, a spatial resolution of approximately 0.5 {\AA}, and a beam current of 101 pA. The transmitted high-angle scattered electrons were collected by an annular detector with inner and outer collection angles of 38 and 200 mrad, respectively. EDS analysis was carried out using an Ultra-X ray detector with a total solid-angle collection of approximately 4.45 sr. Spatially resolved elemental maps were obtained by selecting the characteristic X-ray lines energy for each element: Bi (M-line at 2.580 keV) and Pt (M-line at 2.122 keV). The total acquisition time was 2.9 minutes with a dwell time of 15 ms per pixel to ensure a sufficient signal-to-noise ratio. Careful consideration of acquisition parameters was implemented to minimize sample drift, following established approaches for the atom-scale characterization of metallic quantum materials \cite{sharma2024,babafemi2025}. 

We performed the magnetization measurements using a Quantum Design MPMS XL SQUID magnetometer equipped with an iQuantum He3 insert, which allowed us to measure down to 0.5 K in fields up to 7~T.  For this, a sample was shaped into a sphere by frictional grinding in an abrasive-walled box by maintaining controlled airflow to round the edges of the irregular shape. For the low-temperature transport measurements, we employed the linear four-probe method. Both electrical transport and heat capacity measurements were conducted via Quantum Design's PPMS with an integrated dilution refrigerator option, which allowed measurements down to 50 mK. 

The $\mu$SR measurements were performed at the M15 beamline at TRIUMF, Vancouver, where we used the DR (dilution refrigerator) spectrometer equipped with a 5~T magnet. The samples were held in place on the sample holder with copper grease and secured using a thin silver foil, which does not stop muons.   We performed the measurements with crystals mounted with their $c$-axis along the beam momentum. We measured the $\mu$SR spectra in zero, longitudinal, and transverse field geometries. In zero and longitudinal-field geometry, muons are planted in the sample with their spins antiparallel to their momentum, and backward and forward positron detectors are used to detect the decay positrons. As the superconducting magnet in the DR spectrometer creates the magnetic field parallel to the beam axis, the transverse field measurements were performed in spin-rotated mode, which involves rotating the muon spins in a vertical direction using perpendicular electric and magnetic fields.  In this case, we used left and right positron detectors to record the asymmetry spectra in TF-$\mu$SR. For the ZF-$\mu$SR measurements, a zero-field environment was established following the process described in ref. \cite{Morris2003} \\

%%%%%%%%%%%%%%%%%%%%%%%%%%%%%%%%%%%%%%%%%%%%%%%%%%%%%%%%%%%%%%%%%%%%%%%%%%%%%%%%%%%%%%%%%%
%%------------------FigureBegin-----------------------%%
\begin{figure*}[]
\begin{center}
\includegraphics[scale=0.60]{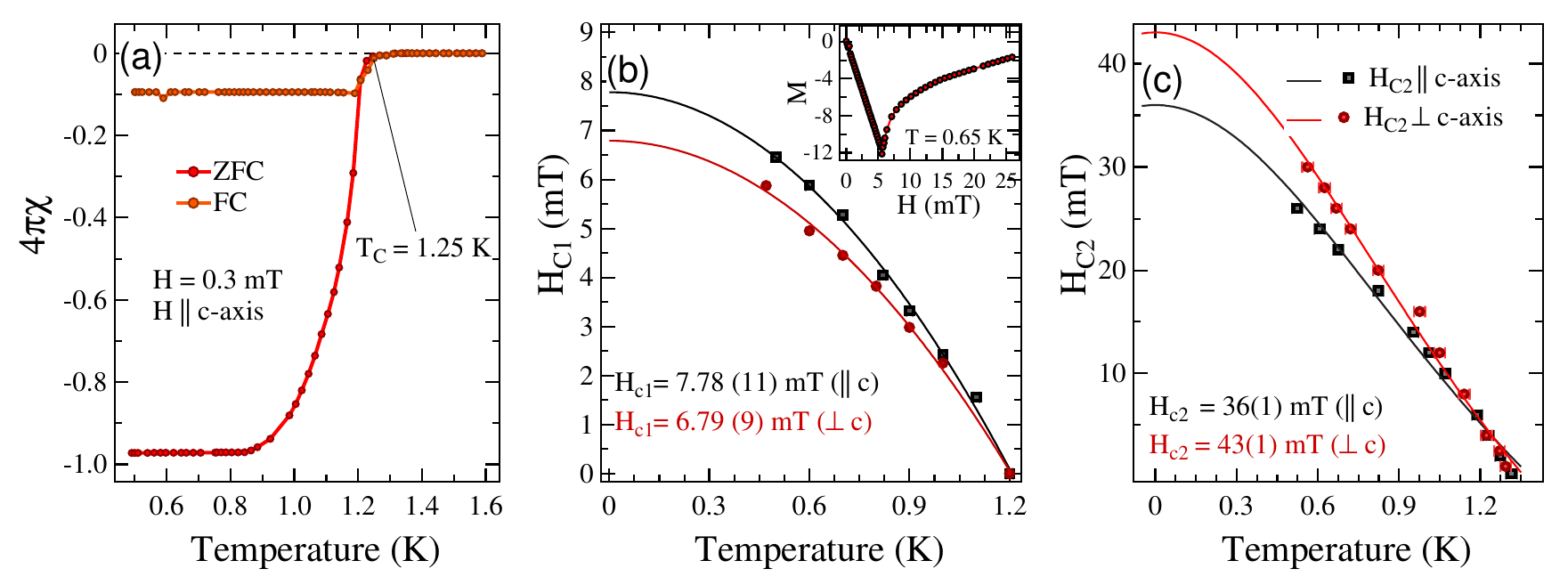}
\caption{(a) ZFC and FC plot showing T$_c$ at 1.25(4) K (b) Temperature dependence of H$_{c1}$ and  Ginzburg-Landau fits along and perpendicular to the c axis of a hexagonal crystal measured on a sphere. The inset shows a representative magnetization (M) versus field (H) plot, which is used to estimate the $H_{c1}$  (c) Temperature dependence of H$_{c2}$ along and perpendicular to the $c$-axis of BiPt crystal. \label{fig2} } 
\end{center}
\end{figure*}
%%------------------FigureEnd-----------------------%%

\section{Structural Characterization}
BiPt crystallizes in a hexagonal crystal structure with P6$_3$/mmc (194) space group with lattice parameters a = b = 4.31394(6) \AA ~and c = 5.49334(12) \AA, $\alpha= \beta = 90\degree$ and $\gamma = 120\degree$ that were determined from powder XRD measurements at room temperature shown in Fig. \ref{fig1}(a). The crystallinity of the Bridgeman-grown single crystal was determined using Laue X-ray diffraction and neutron pole figures. Because of the less penetrative nature of the X-rays, X-ray Laue is ideal for probing the outer surface (few micrometers),  depending on the wavelength of the source, density of material etc. The six-fold pattern shown in Fig. \ref{fig1}(a)[Inset] corresponds to <001> plane of the crystal. Neutrons, in contrast, are more penetrative and therefore can probe the bulk nature of the specimen. The three peaks in the neutron pole figure image in Fig. \ref{fig1}(d) indicate that our BiPt sample consists of a single crystal. We utilized HAADF-STEM and EDS experiments to determine the microscopic arrangements of the atoms, as well as their relative concentration. The EDS data present evidence for an equal concentration (50:50 $\pm$ 3.2) of Bi and Pt. The HAADF-STEM image in Fig. \ref{fig1}(d) shows the contrast between the two types of atoms, where Pt columns occupy the center of hexagonal columns formed by Bi atoms.

%%%%%%%%%%%%%%%%%%%%%%%%%%%%%%%%%%%%%%%%%%%%%%%%%%%%%%%%%%%%%%%%%%%%%%%%%%%%%%%%%%%%%%%%%%%%%%%%%%%%%%%%
\section{Magnetization}

BiPt undergoes a bulk superconducting transition at $\approx$1.2 K. Fig. \ref{fig2} (a) shows the onset of diamagnetism at $\approx$1.2 K  with the application of 0.3~mT field along [001] direction. BiPt displays anisotropic superconducting properties, which can be attributed to its hexagonal crystal structure. BiPt exhibits complete field expulsion up to a lower critical field $H_{c1}$, characterized by linear magnetization versus field behavior at constant temperature,  as shown in Fig. \ref{fig2} 
 [Inset].  Above $H_{c1}$, the field penetrates and the sample enters a vortex state, corresponding to type-II behavior.  We determined $H_{c1}$ at various temperatures from magnetization versus temperature curves. The measurements were performed on a sphere, and therefore $H_{c1}$ was corrected for demagnetizing effects using $H_{c1}=(3/2)H_{c1(applied)}$, here $H_{c1(applied)}$ is the lower critical field without considering demagnetizing effects. We fitted the data with the Ginzburg-Landau equation to estimate $H_{c1}$(0), as shown in Fig. \ref{fig2}[b]. Similarly, the temperature dependence of the upper critical field, $H_{c2}$, was fitted with the Ginzburg-Landau equation to estimate $H_{c2}(0)$. We calculated the coherence length, $\xi$ ( 95.61 nm || c and 87.48 nm $\perp$c)  using $\xi=\sqrt{\phi_0/2\pi H_{c2}(0)}$. The penetration depth, $\lambda_{GL}$, can also be calculated using the Ginzburg-Landau equation: $ H_{c1} = \frac{\phi_0}{2\pi\lambda_{GL}^2}\log(\frac{\lambda_{GL}}{\xi})$.
However, the above equation only yields solutions when $\lambda_{GL} >> \xi$; therefore, we used equation 4 from reference \cite{Brandt2003} to estimate the Ginzburg-Landau parameter $\kappa$  (1.86 || c,  2.27 $\perp$c). Using this $\kappa = \lambda_{GL}/\xi$, we can evaluate $\lambda_{GL}$, 178.6(2.5) nm in the $ab$ plane and 198.4(2.3) nm in the plane perpendicular to the $ab$ plane  ).      

%%--------------------FigureBegin-----------------------------%%
\begin{figure*}[]
\begin{center}
\includegraphics[scale=0.6]{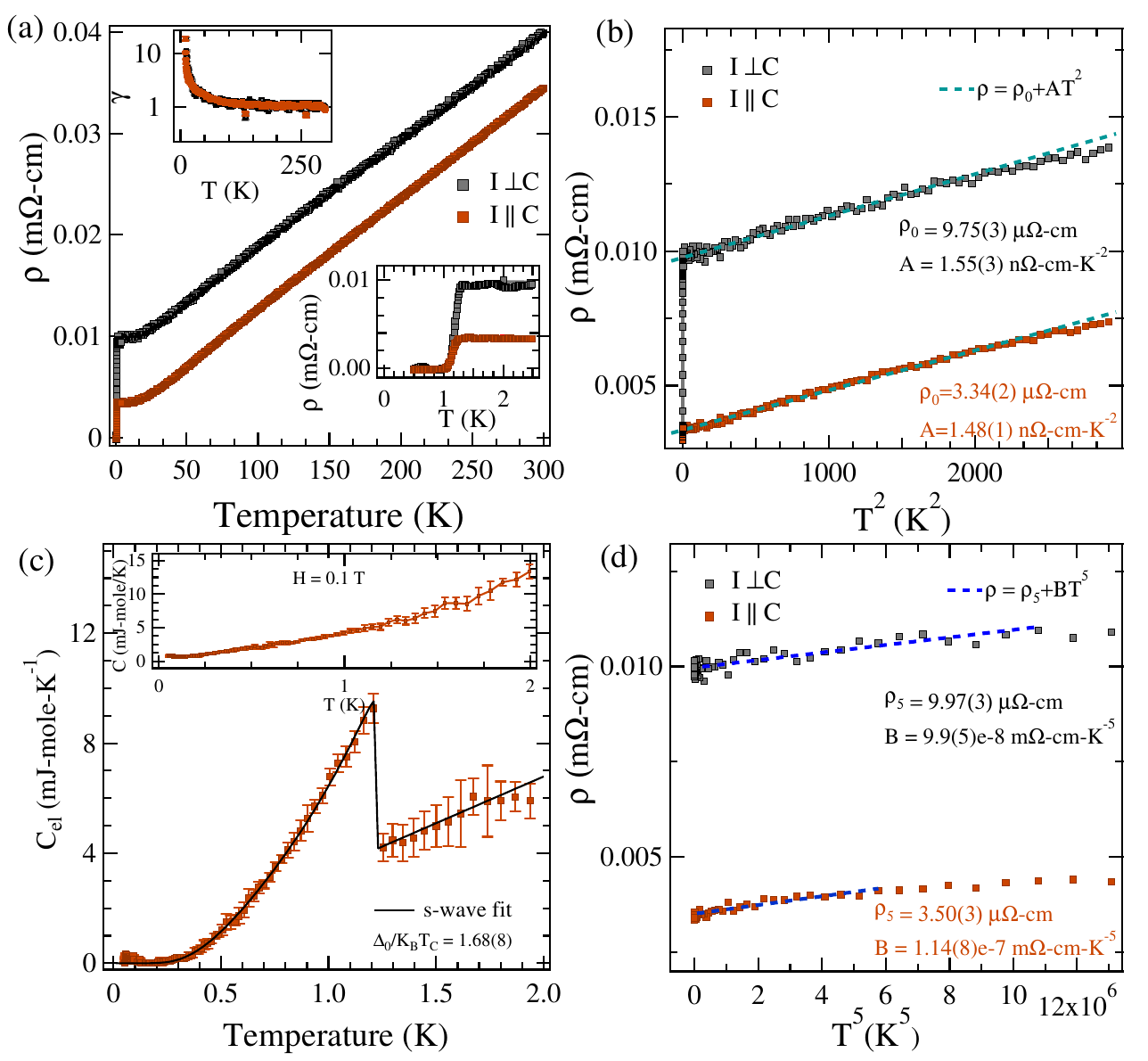}
\caption{ Resistivity and heat capacity data. (a) Resistivity with current perpendicular and parallel to the $c$-axis is shown.[inset-top] The exponent of the inelastic resistivity ($\rho=\rho_0 + A_{\gamma}T^{\gamma}$) determined from the logarithmic derivative, $\gamma = \frac{\delta[\ln(\rho-\rho_0)]}{\delta[\ln(T)]}$. [inset-bottom] The expanded low-temperature region displays a transition from a resistive to a superconducting state at about 1.2 K (b) The resistivity fits to a quadratic behaviour up to 35 K. (d) The resistivity can also be described by $\rho = \rho_5 + BT^5$ up to 25 K, suggesting that the exponent of the resistivity can not be determined accurately at low termpatures.   (c) Electronic specific heat as a function of temperature. [Inset] The specific heat in field of 0.1 T shows the absence of the superconducting transition. \label{ResistivityAndSH}  } 
\end{center}
\end{figure*}
%%----------------------FigureEnd------------------------------%%

%%%%%%%%%%%%%%%%%%%%%%%%%%%%%%%%%%%%%%%%%%%%%%%%%%%%%%%%%%%%%%%%%%%%%%%%%%%%%%%%%%%%%%%
\section{Heat capacity}
Evidence of the bulk nature of superconductivity is apparent through a distinct anomaly observed in the specific heat data at zero field around $T_{c}$ = 1.23 K. When the measurement was conducted at 0.1 T, which exceeds the upper critical field, superconductivity was completely suppressed. The specific heat data, depicted in the inset of Fig. \ref{ResistivityAndSH}(c) as C/T vs $T^{2}$, demonstrates that the normal-state specific heat in the low-temperature range can be accurately described by the relation $\frac{C}{T}=\gamma_{n}+\beta_{3} T^{2} + \beta_{5} T^{4}$. Here, $\gamma_{n}$ accounts for the electronic contribution, while $\beta_{3}$ and $\beta_{5}$ represent the phononic contribution. Through fitting, the following values were obtained: $ \gamma_{n}$ = 2.76 $\pm$ 0.51 mJ/mol K$^{2}$, $\beta_{3}$ = 0.99 $\pm$ 0.15 mJ/mol K$^{4}$, and $\beta_{5}$ = 0.006 $\pm$ 0.009 mJ/mol K$^{6}$. These values enable the estimation of various parameters that characterize the superconducting state. 

By substituting the number of atoms per formula unit, $N$ = 2, into equation \ref{eq:thetaD}, the Debye temperature of the sample ($\theta_{D}$) can be calculated as $\theta_{D}$ = 157.6(8.0) K using the relation \cite{kittel2004introduction}: 

\begin{equation}
\theta_{D}= \left(\frac{12\pi^{4}RN}{5\beta_{3}}\right)^{\frac{1}{3}}
\label{eq:thetaD}
\end{equation}

Furthermore, the obtained value of the Debye temperature can be utilized to estimate the electron-phonon coupling strength in this material using the McMillan relation \cite{McMillen1968}:

\begin{equation}
\lambda_{e-ph} = \frac{1.04+\mu^{}\ln(\theta_{D}/1.45T_{c})}{(1-0.62\mu^{})\ln(\theta_{D}/1.45T_{c})-1.04 }
\label{eq:lambdaEPh}
\end{equation}

Here, $ \mu^{*}$ represents the Coulomb pseudopotential, typically around 0.13 for intermetallic superconductors. The calculated value, $\lambda_{e-ph}$ = 0.52643, indicates the weakly coupled nature of this material.

The value of the specific heat jump for the sample around the superconducting transition, $\Delta C_{el}/\gamma_{n}T_{C}$ = 1.12 $ \pm $ 0.07, is lower than the BCS value of 1.43. The electronic contribution to the specific heat can be calculated by subtracting the phononic contribution from the total heat capacity using the relation $C_{el}=C-(\beta_{3} T^{3} + \beta_{5} T^{5})$.  We find an exponential increase in $C_{el}$, which is consistent with an isotropic superconducting gap. To further elucidate this, we tried to fit the  specific heat data to the conventional BCS form  \cite{padamsee1973quasiparticle}:
\begin{equation}
\frac{S}{\gamma_{n}T_{C}} = -\frac{6}{\pi^2}\left(\frac{\Delta_{0}}{k_{B}T_{C}}\right)\int_{0}^{\infty}[ \textit{f}\ln(f)+(1-f)\ln(1-f)]dy \\
\label{eq:swave}
\end{equation}

\noindent here  $\textit{f}$($\xi$) = [exp($\textit{E}$($\xi$)/$k_{B}T$)+1]$^{-1}$ is the Fermi function, $\textit{E}$($\xi$) = $\sqrt{\xi^{2}+\Delta^{2}(t)}$, where E($ \xi $) is the energy of the normal electrons measured relative to Fermi energy, $\textit{y}$ = $\xi/\Delta(0)$, $\mathit{t = T/T_{C}}$ and $\Delta(t)$ = tanh[1.82(1.018(($\mathit{1/t}$)-1))$^{0.51}$] is the BCS approximation for the temperature dependence of energy gap. Differentiating  Eq. \ref{eq:swave} gives the total electronic specific heat as:

\begin{equation}
\frac{C_{el}}{\gamma_{n}T_{C}} = t\frac{d(S/\gamma_{n}T_{C})}{dt}. \\
\label{eq:Cel}
\end{equation}

The solid black curve in Fig. \ref{ResistivityAndSH} (c) shows a fit of the data to this model, with a superconducting gap value of $\Delta_{0}/k_{B}T_{C}$ = 1.68 $\pm$ 0.08,  consistent with the BCS value of 1.73.

%%%%%%%%%%%%%%%%%%%%%%%%%%%%%%%%%%%%%%%%%%%%%%%%%%%%%%%%%%%%%%%%%%%%%%%%%%%%
\section{Electrical Resistivity}
Resistivity measurements were performed in a linear contact geometry with current flowing along and perpendicular to the $c$-axis. BiPt displays anisotropic resistivity, as expected for a compound with a hexagonal crystal structure.  The temperature dependence of the resistivity can be fitted with a general exponential function ($\rho=\rho_0 + A_{\gamma}T^{\gamma}$)). We can take the logarithmic derivative of the resistivity to determine the exponent of the resistivity $\gamma = \frac{\delta[\ln(\rho-\rho_0)]}{\delta[\ln(T)]}$ \cite{TransportPRX2022}. $\gamma$ for two configurations of resistivity for BiPt is shown in Fig. \ref{ResistivityAndSH}(a) [inset-top], which is undefined below 10 K as the resistivity becomes constant and approaches one as the temperature increases above 75 K. The intermediate temperature range resistivity can be approximated with an exponent  $\gamma$=2, consistant with Fermi-liquid behavior as shown in Fig. \ref{ResistivityAndSH}(b). However, the resistivity data can also be reasonably fitted using $\gamma$=5 as shown in \ref{ResistivityAndSH}(d). This suggests that the resistivity exponent at low temperatures can not be uniquely determined. Therefore, we conclude that the resistivity is linear at high temperatures and super-linear at low temperatures, consistent with Bloch-Gr\"uneisen model \cite{Grimvall1981TheEI}.
%We fitted the normal state resistivity data with Bloch-Gr\"uneisen model \cite{Grimvall1981TheEI}
%\begin{equation}
%    \rho(T) = \rho_0 + B(\frac{T}{\theta_D})^{n'}\int_{0}^{\theta_D/T}\frac{x^{n'}}{(e^x - 1)(1-e^{-x})}dx
%\end{equation}

%for simple metals ($n'$=5), to extract the values of Debye temperature ($\theta_D$) and temperature-independent residual resistivity term, $\rho_0$, as shown in Fig. \ref{ResistivityAndSH} (a).  The Debye temperatures ($\theta_D$) obtained from the fit are  144(2) and 149(1) K along and perpendicular to the $c$-axis, respectively, consistent with the heat capacity-derived value (157.6(8.0) K). 
The residual resistivity ratio ($\rho$(300)/$\rho$(2)) is 4 and 9.9 along and perpendicular to the $c$-axis, respectively, suggesting low disorder in the measured single crystals. The residual resistivity terms ($\rho_0$) are 9.75(3) and 334(2) $\mu \Omega$-cm along and perpendicular to the $c$-axis, indicative of the anisotropic nature of electronic properties.  Hall measurements reveal a high carrier density of n =$1.68(31)\times10^{28} /m^3$, consistent with metallic behavior.
The Sommerfeld coefficient $\gamma_n$, obtained from heat capacity measurements, can be expressed as:
\begin{equation}
    \gamma_n = \left( \frac{\pi}{3} \right)^{2/3} \frac{k_B^2 m^* V_{f.u.}N_A n^{1/3}}{\hbar^2},
\end{equation}
here \( k_B \) is the Boltzmann constant, \( V_{f.u.} \) is the volume of the formula unit, and \( N_A \) is Avogadro's number. This equation provides a way to estimate the quasiparticle effective mass, $m^*$. Additionally, we can also determine the Fermi velocity, $\nu_F$, using $n$ and $m^*$
\begin{equation}
    n = \frac{m^{*3} \nu_F^3}{3 \pi^2 \hbar^3}.
\end{equation}
We can estimate the mean free path \( l_e \) from the residual resistivity, \( \rho_0 \), through the relation:
\begin{equation}
    l_e = \frac{3\pi^2 \hbar^3}{e^2 \rho_0 m^{*2} \nu_F^2}
\end{equation}
 We can estimate an effective London penetration depth \( \lambda_L \) from \( n \) via \cite{ashcroft1976solid}
\begin{equation}
    \lambda_L = \sqrt{\frac{m^*}{\mu_0 n e^2}},
\end{equation}
where \( \mu_0 \) is the magnetic permeability of free space. The London penetration depth, $\lambda_L$ = 65(19)~nm, obtained is small compared to $\lambda_{GL}$ estimated from the magnetization measurements. This could be due to the fact that the London model is a simplistic treatment of a single-band isotropic material, unlike BiPt, which is a multiband anisotropic system \cite{sharma2024}.   When impurity scattering is considered, the Ginzburg-Landau penetration depth, \( \lambda_{GL} \), is related to the London penetration depth as:
\begin{equation}
    \lambda_{GL} = \lambda_L \sqrt{1 + \frac{\xi_0}{l_e}},
\end{equation}
The estimated coherence lengths $\xi_0$  (127.74 nm, 62.4 nm ) exceed the mean free paths $l$ (55.4(11.7) nm, 19.5(4.1) nm), which indicates that BiPt is a dirty limit superconductor. The values of the parameters calculated from the above equations are listed in Table \ref{tab:sc}.

%%%%%%%%%%%%%%%%%%%%%%%%%%%%%%%%%%%%%%%%%%%%%%%%%%%%%%%%%%%%%%%%%%%%%%%%%%%%%%%%
\section{Transverse field $\mu$SR}
\begin{figure*}[btp]
\begin{center}
\includegraphics[scale=0.60]{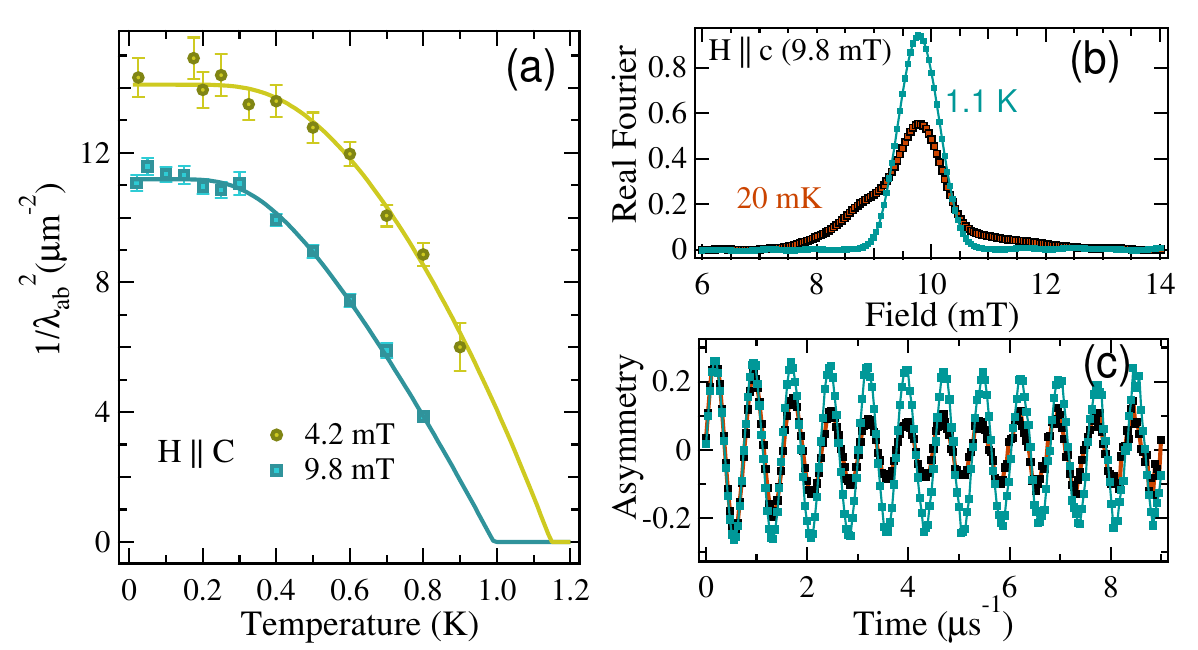}
\caption{Results of the transverse field $\mu$SR measurements performed on single crystalline BiPt with 4.2 mT and 9.8 mT fields along the $c$-axis. (a) The temperature dependence of the inverse penetration depth square in the ab-plane and its $s$-wave fit when the fields were applied along the $c$-axis. The penetration depth was obtained from the fits of TF data with the iterative GL model. (b) The Fourier transform of the TF-SR data shows the comparison field distribution in the ab-plane when the out-of-plane field is applied below (blue) and above (dark yellow) the superconducting transition (c) The asymmetry spectra show increased relaxation upon cooling below the superconducting transition due to the formation of flux line lattice (FLL).       \label{muSR}  } 
\end{center}
\end{figure*}

We performed transverse field $\mu$SR measurements to investigate the geometry and magnitude of the superconducting gap. In order to measure the TF-$\mu$SR spectra, we applied axial fields to the crystalline plates of BiPt such that their out-of-plane axis was aligned along the $c$-axis. A representative  TF spectra and their Fourier transforms are shown in Fig. \ref{muSR}(b-c). As the material enters the superconducting state in the presence of a field, the development of the flux-line lattice  causes an inhomogeneous field distribution, leading to increased relaxation of the asymmetry signal. We used a two-component sinusoidally oscillating relaxing function,
\begin{equation}
\begin{split}
G_{\mathrm{TF}}(t) = a_s \;\mathrm{exp}\left(-(\sigma_{sc}^2+\sigma^2_n) t^2/2\right)\mathrm{cos}(\omega_{s}t+\phi)\\+\;a_b\;\mathrm{exp}\left(-\psi t\right)\mathrm{cos}(\omega_{b}t+\phi),
\label{eqTF}
\end{split}
\end{equation}

\noindent to fit the transverse field data. Here, $a_s$ and $a_b$ are sample and background asymmetries, respectively.  $\sigma$ is the Gaussian relaxation rate of the sample, $\phi$ is the initial phase of the muons, and $\psi$ is the exponential relaxation rate due to the background. $\omega_s$, and $\omega_b$ are the average frequencies of the muon precession in the sample and background, respectively. The field experienced by the muon is proportional to the frequency as $\omega = H\gamma_{\mu}$, where $H$ is the field and $\gamma_{\mu}$ ($2\pi\times135.53MHz/T$) is the gyromagnetic ratio of the muon. 
 $\sigma_n$ is the temperature-independent contribution to the sample relaxation rate originating from the nuclear moments of the constituent atoms, whereas $\sigma_{sc}$ is the relaxation due to the underlying vortex lattice. 

The TF-$\mu$SR data were fit using Brandt's iterative Ginzburg-Landau (GL) model \cite{BrandtIGL, SonierV}. This model generates the field distribution due to the vortex lattice, $B(\textbf{r})$, for all values of $\kappa$, and applied field $b$. The sample term in Eq.(\ref{eqTF}) was replaced with 
\begin{equation}
    A_s(t) = a_se^{-(\sigma_n^2+\sigma^2_{dis})t^2} \sum_\textbf{r}\cos[\gamma_{\mu}B(\textbf{r})t+\phi]
\end{equation}
\noindent using $B(\textbf{r})$ from the set of equations in ref. \cite{SonierV}; the sum is over the ideal vortex lattice in the real space.  $\sigma_{dis}$ accounts for further broadening due to deviations from the ideal hexagonal vortex lattice arising from the frozen disorder.
The iterative GL model produced good fits to the data as a result of a well-formed flux line lattice in the $ab$ plane. This is evident from the high-field tail seen in the Fourier transform (Fig. \ref{muSR}(b)) of the asymmetry spectra(Fig. \ref{muSR}(c)).  We obtained the absolute values of the penetration depth from the fits and used them to determine the temperature dependence of the superfluid density ($\propto$ $1/\lambda^2$).  The temperature dependence of the 4.2 mT and 9.8 mT temperature scans shows flattening of the superfluid density at low temperature, as expected from a conventional $s$-wave superconductor due to an isotropic superconducting gap. The difference between the zero-temperature penetration depth values for the two fields indicates a field dependence. Comparing these with the values obtained from the magnetization measurements [Table: \ref{tab:sc}], we observe a monotonic increase of the penetration depth with the field expected from a single-band type-II superconductor \cite{SonierV}. The difference arises from the fact that the magnetization measurements of penetration depth are done in the Meissner state (low fields), whereas the $\mu$SR results are from the vortex state.  Similarly, $\kappa$ also exhibits field dependence. 

We can estimate the superconducting gap from the temperature dependence of penetration depth using London's approximation for a BCS superconductor in the dirty limit, 
\begin{equation}    
\frac{\lambda^{-2}(T)}{\lambda^{-2}(0)} = \frac{\Delta(T)}{\Delta(0)} \tanh\left( \frac{\Delta(T)}{2 k_B T} \right).
\end{equation}

For completeness, we  fit the results to the clean limit as well, 
	\begin{equation}
     \frac{\lambda^{-2}(T)}{\lambda^{-2}(0)} = 1+2\int_{|\Delta(T)|}^{\infty}\left(\frac{\delta f}{\delta E}\right)\frac{E dE}{\sqrt{E^{2}-\Delta^{2}(T)}}  ,
    \label{eqnclean}
    \end{equation}
here $f = [1+\exp(E/k_{B}T)]^{-1}$ is the Fermi function, and
\begin{equation}
   \Delta(T) = \Delta(0)\tanh[1.82(1.018((\mathit{T_{c}/T})-1))^{0.51}] 
\end{equation}
 is the temperature dependence of the energy gap in the BCS approximation \cite{gapequation}, where $\Delta(0)$ is the superconducting gap value at zero temperature.  The values of the superconducting gaps in the $ab$-plane in the two limits, as well as their field dependence, are listed in Table \ref{tab:sc}.  The superconducting gaps in the dirty limit, is smaller than the BCS value of 1.76, and  appears to change with  field, (1.32(5) versus 1.53(22)), which may indicate a multi-band nature, although the change is within error bars. The multi-band nature is expected in BiPt due multiple bands crossing the Fermi level \cite{sharma2024} in these family of compounds.

\begin{table*}[]
    \centering
    \begin{tabular}{|l|l|l|l|l|}
    \hline
         &\multicolumn{3}{c|}{in ab-plane}  &out of ab-plane \\ \hline %H||c & H perpendicular to c%
        &\multicolumn{2}{c|}{$\mu$SR in field of} &other &other  \\
          &9.2 mT& 4.2mT&  techniques & techniques   \\ %H||c & H perpendicular to c%
         \hline
         T$_{c,clean}$(K)& 0.992(16)&1.15(6)& &\\
         \hline
         $\Delta_{0,clean}$ (meV)&0.1426(30)&0.187(8) && \\
         \hline
         $\Delta_{0,\mathrm{clean}}/(k_B T_c)$ &1.67(4)&1.32(5)  && \\
         \hline
         T$_{c,\mathrm{dirty}}$(K)& 0.973(11)&1.15(7) &&\\
         \hline
         $\Delta_{0,dirty}$ (meV)&0.111(4)&0.152(19) && \\
         \hline
         $\Delta_{0,dirty}/(k_B T_c )$&1.32(5)&1.53(22) &&\\
         \hline
         $\Delta_{0,HC}/(k_B T_c )$ &&&\multicolumn{2}{c|}{1.68(8)}  \\ 
         \hline
        $n(m^{-3})$ &&&\multicolumn{2}{c|}{$1.68(31)\times10^{28}$}  \\ 
         \hline
          $\gamma_{n}(mJ/mol K^2)$ &&&\multicolumn{2}{c|}{2.67(51)}  \\ 
         \hline
         H$_{C1}$(mT)&&&7.78(11)  & 6.79(9) \\
         \hline
         H$_{C2}$(mT)&&&36(1)&   43(1) \\ 
         \hline
         $\lambda$(nm)& 298.5(5.6)&264.5(4.8)&178.6(2.4) &199.4(2.3)  \\
         \hline
%         $\lambda_{GL}$(nm)&&&163.47  & 217.84 &\\
%         \hline
         $\xi_{mag}$(nm)&& & 95.6(1.3)& 87.4(1.0)  \\
         \hline
%         $\kappa_{mag}$&&&2.50 & 2.60 &\\ 
%         \hline
         $\kappa$& 5.31(6)&4.4(3)&2.5&2.6  \\ 
         \hline
         $\rho_0$($\mu\Omega$-cm)&&&3.34(2)  & 9.75(3) \\ 
         \hline
         $\theta_D$(K)&&&\multicolumn{2}{c|}{157.6(8.0)} \\ 
         \hline
         $l$(nm)&&&55.4(11.7)  & 19.5(4.1)\\ 
        \hline
        $m^*/m_e$&&& \multicolumn{2}{c|}{2.51(26)} \\ 
        \hline
        $\lambda_L$(nm)&&& \multicolumn{2}{c|}{65(19)} \\ 
        \hline
         $\xi_0$(nm)&&&127.74  &62.4 \\ 
        \hline
         
    \end{tabular}
\caption{Superconducting and normal state parameters table based on $\mu$SR data in an applied field and non-$\mu$SR (resistivity, heat capacity and magnetization) data. }
    \label{tab:sc}
\end{table*}

%%%%%%%%%%%%%%%%%%%%%%%%%%%%%%%%%%%%%%%%%%%%%%%%%%%%%%%%%%%%%%%%%%%%%%%%%%%%%%
\section{Zero-field $\mu$SR}

%%------------------FigureBegin-----------------------%%
\begin{figure}[]
\begin{center}
\includegraphics[scale=0.51]{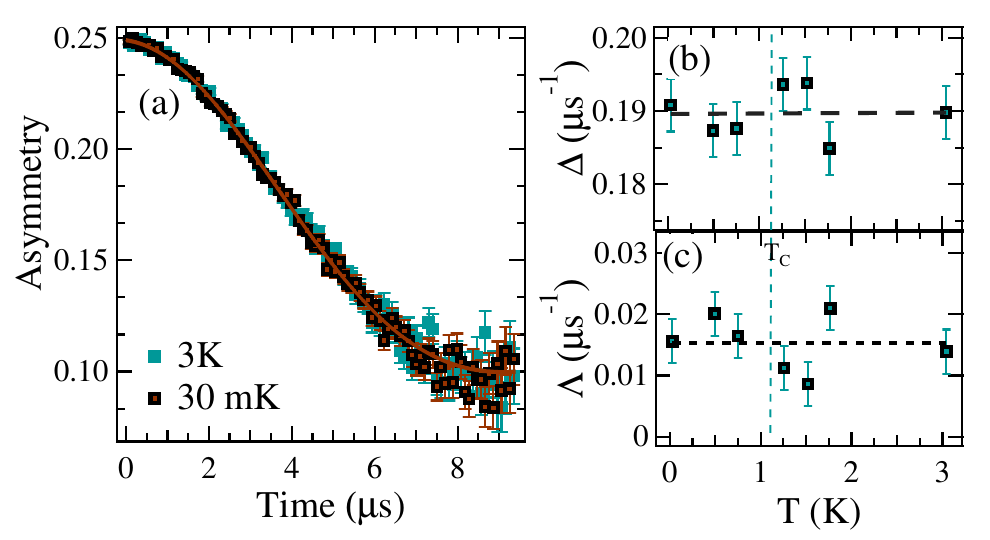}
\caption{ (a) Zero-field $\mu$SR spectra above and below the superconducting transition, where the initial muon polarization was parallel to the $c$-axis. The spectra remain unchanged within the resolution of $\mu$SR, as evident from the relaxation rate parameters (b) $\Lambda$ and (c) $\Delta$, which do not show any systematic increase below T$_c$. This suggests that time-reversal symmetry is preserved in BiPt.  \label{ZF} } 
\end{center}
\end{figure}
%%------------------FigureEnd-----------------------%%

Muon spin relaxation measurements performed in zero field (ZF) can provide conclusive evidence of a time-reversal symmetry (TRS) breaking superconducting state. Fig. \ref{ZF}(a) shows the ZF spectra above and below the transition temperature. When the polarized muons come across random nuclear moments, their polarization evolves over time. 
This muon depolarization can be described by the Gaussian Kubo-Toyabe function \cite{KT1979},
\begin{equation}
 G_{\mathrm{KT}}(t) = \frac{1}{3}+\frac{2}{3}(1-\Delta^{2}t^{2})\mathrm{exp}\left(-\frac{\Delta^{2}t^{2}}{2}\right) ,
 \label{eqn1:zf}
 \end{equation} 
 where $\sigma$ is the width of the nuclear dipolar field experienced by the muons, and $t$ is time.
 
 One can use the following relaxation function to fit the ZF-$\mu$SR spectra of the superconductors
\begin{equation}
 A(t) = A_{s}G_{\mathrm{KT}}(t)\mathrm{exp}(-\Lambda t)+A_{\mathrm{BG}} 
 \label{eqn:tay}
 \end{equation}
 Here, $  A_{BG} $ is the background asymmetry, $  A_{s} $ is the sample asymmetry, and the exponential term accounts for additional relaxation originating from magnetic fields. A superconductor that breaks time-reversal symmetry exhibits an increased relaxation rate below the transition temperature \cite{sharma2023, luke1998}. However, for BiPt, the spectra remain unchanged between normal and superconducting state, as shown in Fig. \ref{ZF}, indicative of a time-reversal preserving superconducting state.  Figures \ref{ZF} (b) and (c) present the temperature dependence of the relaxation rate parameters, $\Lambda$, and $\Delta$ obtained from the fits of the ZF data.

%%%%%%%%%%%%%%%%%%%%%%%%%%%%%%%%%%%%%%%%%%%%%%%%%%%%%%%%%%%%%%%%%%%%%%%%%%%%%%%%%
\section{Discussion and Conclusions }

A recent report on the topological properties of the Bi$_2$Pd$_x$Pt$_{1-x}$ compounds revealed the topological trivial nature of BiPt (x=0) \cite{sharma2024} and that Pd-substituted systems may realize Dirac surface states; the binding energy of the Dirac states is farthest from the Fermi surface in the case of BiPt. We studied the normal and superconducting state properties of this material, which exhibits anisotropic properties due to its hexagonal crystal lattice.  In the normal state, BiPt is a metal with hole-like carriers.   BiPt is a weak type-II superconductor with Ginzburg-Landau parameter $\kappa_{GL}$ exceeding $1/\sqrt{2}$ (1.86 and 2.27), but that is smaller than most type-II superconductors. Our $\mu$SR measurements point to the field-sensitive nature of the penetration depth. BiPt exhibits a nodeless and anisotropic superconducting gap in the weak coupling regime, likely due to its multi-band nature. The superconducting state preserves time-reversal symmetry. Thus, we conclude that BiPt is a conventional $s$-wave superconductor with topologically trivial properties. These results on BiPt are in stark contrast to those of similar systems, such as $t$-PtBi$_2$, which is also a hexagonal material with topologically non-trivial surface superconductivity\cite{gao2018, schimmel2024}. Isostructural $\gamma$-BiPd is also a topologically non-trivial superconductor \cite{chiang2023,sharma2024}. Therefore, our work will serve as a valuable reference, as BiPt is a suitable comparison system with other similar Bi/Pt/Pd systems, which exhibit a diverse range of exotic properties.

%%%%%%%%%%%%%%%%%%%%%%%%%%%%%%%%%%%%%%%%%%%%%%%%%%%%%%%%%%%%%%%%%%%%%%%%%%%%%%
\section{Acknowledgments}
Work at McMaster was supported by the Natural Sciences and Engineering Research of Council of Canada. M.J.L. and B.S.A. acknowledge the financial support of the Natural Sciences and Engineering Research Council of Canada (NSERC) under the Discovery Grant Program. We also thank the Canadian Centre for Electron Microscopy (CCEM) for providing access to electron microscopy facilities. Use of the MAD beamline at the McMaster Nuclear Reactor is supported by McMaster University and the Canada Foundation for Innovation. The work at the University of Minnesota was funded by the U.S. Department of Energy through the University of Minnesota,  Center for Quantum Materials, under Grant No. DE-SC-0016371.

\section{Data Availability}
The $\mu$SR data supporting the findings of this article are publicly available at \cite{muSRdata}. The rest of the data supporting the findings of this article is available at reasonable request from the authors.

\bibliography{BiPt}

@article{McMillen1968,
  title = {Transition Temperature of Strong-Coupled Superconductors},
  author = {McMillan, W. L.},
  journal = {Phys. Rev.},
  volume = {167},
  issue = {2},
  pages = {331--344},
  numpages = {0},
  year = {1968},
  month = {Mar},
  publisher = {American Physical Society},
  doi = {10.1103/PhysRev.167.331},
  url = {https://link.aps.org/doi/10.1103/PhysRev.167.331}
}

@article{padamsee1973quasiparticle,
  title={Quasiparticle phenomenology for thermodynamics of strong-coupling superconductors},
  author={Padamsee, H and Neighbor, JE and Shiffman, CA},
  journal={Journal of Low Temperature Physics},
  volume={12},
  pages={387--411},
  year={1973},
  publisher={Springer}
}

@article{Brandt2003,
  title = {Properties of the ideal Ginzburg-Landau vortex lattice},
  author = {Brandt, Ernst Helmut},
  journal = {Phys. Rev. B},
  volume = {68},
  issue = {5},
  pages = {054506},
  numpages = {11},
  year = {2003},
  month = {Aug},
  publisher = {American Physical Society},
  doi = {10.1103/PhysRevB.68.054506},
  url = {https://link.aps.org/doi/10.1103/PhysRevB.68.054506}
}

@article{Morris2003,
title = {A method of achieving accurate zero-field conditions using muonium},
journal = {Physica B: Condensed Matter},
volume = {326},
number = {1},
pages = {252-254},
year = {2003},
issn = {0921-4526},
doi = {https://doi.org/10.1016/S0921-4526(02)01618-6},
url = {https://www.sciencedirect.com/science/article/pii/S0921452602016186},
author = {G.D. Morris and R.H. Heffner}
}

@Article{Sakano2015,
author={Sakano, M.
and Okawa, K.
and Kanou, M.
and Sanjo, H.
and Okuda, T.
and Sasagawa, T.
and Ishizaka, K.},
title={Topologically protected surface states in a centrosymmetric superconductor {$\beta$-PdBi$_2$}},
journal={Nature Communications},
year={2015},
month={Oct},
day={13},
volume={6},
number={1},
pages={8595},
issn={2041-1723},
doi={10.1038/ncomms9595},
url={https://doi.org/10.1038/ncomms9595}
}

@article{gapequation,
author = {Carrington, A. and Manzano, F.},
doi = {10.1016/S0921-4534(02)02319-5},
issn = {09214534},
journal = {Physica C: Superconductivity and its Applications},
month = {mar},
number = {1-2},
pages = {205--214},
publisher = {North-Holland},
title = {{Magnetic penetration depth of MgB2}},
volume = {385},
year = {2003}
}

@article{Fu_Kane_2008,
  title = {Superconducting Proximity Effect and Majorana Fermions at the Surface of a Topological Insulator},
  author = {Fu, Liang and Kane, C. L.},
  journal = {Phys. Rev. Lett.},
  volume = {100},
  issue = {9},
  pages = {096407},
  numpages = {4},
  year = {2008},
  month = {Mar},
  publisher = {American Physical Society},
  doi = {10.1103/PhysRevLett.100.096407},
  url = {https://link.aps.org/doi/10.1103/PhysRevLett.100.096407}
}

@article{Benia_Wahl_2016,
  title = {Observation of Dirac surface states in the noncentrosymmetric superconductor {BiPd}},
  author = {Benia, H. M. and Rampi, E. and Trainer, C. and Yim, C. M. and Maldonado, A. and Peets, D. C. and St\"ohr, A. and Starke, U. and Kern, K. and Yaresko, A. and Levy, G. and Damascelli, A. and Ast, C. R. and Schnyder, A. P. and Wahl, P.},
  journal = {Phys. Rev. B},
  volume = {94},
  issue = {12},
  pages = {121407},
  numpages = {5},
  year = {2016},
  month = {Sep},
  publisher = {American Physical Society},
  doi = {10.1103/PhysRevB.94.121407},
  url = {https://link.aps.org/doi/10.1103/PhysRevB.94.121407}
}

@article{sun2015dirac,
  title={Dirac surface states and nature of superconductivity in noncentrosymmetric {BiPd}},
  author={Sun, Zhixiang and Enayat, Mostafa and Maldonado, Ana and Lithgow, Calum and Yelland, Ed and Peets, Darren C and Yaresko, Alexander and Schnyder, Andreas P and Wahl, Peter},
  journal={Nature Communications},
  volume={6},
  number={1},
  pages={6633},
  year={2015},
  publisher={Nature Publishing Group UK London},
   url={https://doi.org/10.1038/ncomms7633}
}

@article{Lv_Xue_2017,
title = {Experimental signature of topological superconductivity and Majorana zero modes on {$\beta$-Bi$_2$Pd} thin films},
journal = {Science Bulletin},
volume = {62},
number = {12},
pages = {852-856},
year = {2017},
issn = {2095-9273},
doi = {https://doi.org/10.1016/j.scib.2017.05.008},
url = {https://www.sciencedirect.com/science/article/pii/S2095927317302487},
author = {Yan-Feng Lv and Wen-Lin Wang and Yi-Min Zhang and Hao Ding and Wei Li and Lili Wang and Ke He and Can-Li Song and Xu-Cun Ma and Qi-Kun Xue},
}

@article{sharma2023,
  title = {Evidence for nonunitary triplet-pairing superconductivity in noncentrosymmetric {TaRuSi} and comparison with isostructural {TaReSi}},
  author = {Sharma, S. and K. P., Sajilesh and Richards, A. D. S. and Gautreau, J. and Pula, M. and Beare, J. and Kojima, K. M. and Yoon, S. and Cai, Y. and Kushwaha, R. K. and Agarwal, T. and S\o{}rensen, E. S. and Singh, R. P. and Luke, G. M.},
  journal = {Phys. Rev. B},
  volume = {108},
  issue = {14},
  pages = {144510},
  numpages = {10},
  year = {2023},
  month = {Oct},
  publisher = {American Physical Society},
  doi = {10.1103/PhysRevB.108.144510},
  url = {https://link.aps.org/doi/10.1103/PhysRevB.108.144510}
}

@article{luke1998,
  title={Time-reversal symmetry-breaking superconductivity in {Sr$_2$RuO$_4$}},
  author={Luke, G Ml and Fudamoto, Y and Kojima, KM and Larkin, MI and Merrin, J and Nachumi, B and Uemura, YJ and Maeno, Y and Mao, ZQ and Mori, Y and Nakamura, H. and Sigrist, M.},
  journal={Nature},
  volume={394},
  number={6693},
  pages={558--561},
  year={1998},
  url = {https://doi.org/10.1038/29038},
  publisher={Nature Publishing Group UK London}
}

@article{Matthias1953,
  title = {Superconducting Compounds of Nonsuperconducting Elements},
  author = {Matthias, B. T.},
  journal = {Phys. Rev.},
  volume = {90},
  issue = {3},
  pages = {487--487},
  numpages = {0},
  year = {1953},
  month = {May},
  publisher = {American Physical Society},
  doi = {10.1103/PhysRev.90.487},
  url = {https://link.aps.org/doi/10.1103/PhysRev.90.487}
}

@article{zhuravlev1959,
  title={On the Superconductivity of the Compound {BiPt}},
  author={Zhuravlev, NN and Stepanova, AA and Zyuzin, NI},
  journal={JETP},
  volume={8},
  pages={1101},
  year={1959},
  url={http://jetp.ras.ru/cgi-bin/dn/e_010_03_0627.pdf}
}

@article{sharma2024,
  title = {Evidence for conventional superconductivity in ${\mathrm{Bi}}_{2}\mathrm{PdPt}$ and prediction of possible topological superconductivity in disorder-free $\ensuremath{\gamma}\text{\ensuremath{-}}\mathrm{BiPd}$},
  author = {Sharma, S. and Richards, A. D. S. and K. P., Sajilesh and Kataria, A. and Agboola, B. S. and Pula, M. and Gautreau, J. and Ghara, A. and Singh, D. and Marik, S. and Dunsiger, S. R. and Lagos, M. J. and Kanigel, A. and S\o{}rensen, E. S. and Singh, R. P. and Luke, G. M.},
  journal = {Phys. Rev. B},
  volume = {109},
  issue = {22},
  pages = {224509},
  numpages = {9},
  year = {2024},
  month = {Jun},
  publisher = {American Physical Society},
  doi = {10.1103/PhysRevB.109.224509},
  url = {https://link.aps.org/doi/10.1103/PhysRevB.109.224509}
}

@article{gao2018,
  title={A possible candidate for triply degenerate point fermions in trigonal layered {PtBi$_2$}},
  author={Gao, Wenshuai and Zhu, Xiangde and Zheng, Fawei and Wu, Min and Zhang, Jinglei and Xi, Chuanying and Zhang, Ping and Zhang, Yuheng and Hao, Ning and Ning, Wei and others},
  journal={Nature communications},
  volume={9},
  number={1},
  pages={3249},
  year={2018},
  publisher={Nature Publishing Group UK London},
  URL ={https://doi.org/10.1038/s41467-018-05730-3}
}

@article{schimmel2024,
  title={Surface superconductivity in the topological Weyl semimetal {t-PtBi$_2$}},
  author={Schimmel, Sebastian and Fasano, Yanina and Hoffmann, Sven and Besproswanny, Julia and Corredor Bohorquez, Laura Teresa and Puig, Joaqu{\'\i}n and Elshalem, Bat-Chen and Kalisky, Beena and Shipunov, Grigory and Baumann, Danny and others},
  journal={Nature Communications},
  volume={15},
  number={1},
  pages={9895},
  year={2024},
  publisher={Nature Publishing Group UK London},
  url={https://doi.org/10.1038/s41467-024-54389-6},
}

@article{wu2020GMR,
  title = {Huge linear magnetoresistance due to open orbits in $\ensuremath{\gamma}\text{\ensuremath{-}}{\mathrm{PtBi}}_{2}$},
  author = {Wu, Beilun and Barrena, V\'{\i}ctor and Suderow, Hermann and Guillam\'on, Isabel},
  journal = {Phys. Rev. Res.},
  volume = {2},
  issue = {2},
  pages = {022042},
  numpages = {6},
  year = {2020},
  month = {May},
  publisher = {American Physical Society},
  doi = {10.1103/PhysRevResearch.2.022042},
  url = {https://link.aps.org/doi/10.1103/PhysRevResearch.2.022042}
}

@article{Gao2017GMR,
  title = {Extremely Large Magnetoresistance in a Topological Semimetal Candidate Pyrite ${\mathrm{PtBi}}_{2}$},
  author = {Gao, Wenshuai and Hao, Ningning and Zheng, Fa-Wei and Ning, Wei and Wu, Min and Zhu, Xiangde and Zheng, Guolin and Zhang, Jinglei and Lu, Jianwei and Zhang, Hongwei and Xi, Chuanying and Yang, Jiyong and Du, Haifeng and Zhang, Ping and Zhang, Yuheng and Tian, Mingliang},
  journal = {Phys. Rev. Lett.},
  volume = {118},
  issue = {25},
  pages = {256601},
  numpages = {5},
  year = {2017},
  month = {Jun},
  publisher = {American Physical Society},
  doi = {10.1103/PhysRevLett.118.256601},
  url = {https://link.aps.org/doi/10.1103/PhysRevLett.118.256601}
}

@article{dimitri2018,
  title = {Dirac state in a centrosymmetric superconductor {$\ensuremath{\alpha}\text{\ensuremath{-}}{\mathrm{PdBi}}_{2}$}},
  author = {Dimitri, Klauss and Hosen, M. Mofazzel and Dhakal, Gyanendra and Choi, Hongchul and Kabir, Firoza and Sims, Christopher and Kaczorowski, Dariusz and Durakiewicz, Tomasz and Zhu, Jian-Xin and Neupane, Madhab},
  journal = {Phys. Rev. B},
  volume = {97},
  issue = {14},
  pages = {144514},
  numpages = {5},
  year = {2018},
  month = {Apr},
  publisher = {American Physical Society},
  doi = {10.1103/PhysRevB.97.144514},
  url = {https://link.aps.org/doi/10.1103/PhysRevB.97.144514}
}

@article{li2019,
  title={Observation of half-quantum flux in the unconventional superconductor {$\beta$-Bi$_2$Pd}},
  author={Li, Yufan and Xu, Xiaoying and Lee, M-H and Chu, M-W and Chien, CL},
  journal={Science},
  volume={366},
  number={6462},
  pages={238--241},
  year={2019},
  publisher={American Association for the Advancement of Science},
  url={https://www.science.org/doi/10.1126/science.aau6539}
}

@article{neupane2016,
  title={Observation of the spin-polarized surface state in a noncentrosymmetric superconductor {BiPd}},
  author={Neupane, Madhab and Alidoust, Nasser and Hosen, M Mofazzel and Zhu, Jian-Xin and Dimitri, Klauss and Xu, Su-Yang and Dhakal, Nagendra and Sankar, Raman and Belopolski, Ilya and Sanchez, Daniel S and others},
  journal={Nature communications},
  volume={7},
  number={1},
  pages={13315},
  year={2016},
  publisher={Nature Publishing Group UK London},
  url={https://doi.org/10.1038/ncomms13315}
}

@article{chiang2023,
  title = {Unequivocal Identification of Spin-Triplet and Spin-Singlet Superconductors with Upper Critical Field and Flux Quantization},
  author = {Chiang, C. C. and Lee, H. C. and Lin, S. C. and Qu, D. and Chu, M. W. and Chen, C. D. and Chien, C. L. and Huang, S. Y.},
  journal = {Phys. Rev. Lett.},
  volume = {131},
  issue = {23},
  pages = {236003},
  numpages = {6},
  year = {2023},
  month = {Dec},
  publisher = {American Physical Society},
  doi = {10.1103/PhysRevLett.131.236003},
  url = {https://link.aps.org/doi/10.1103/PhysRevLett.131.236003}
}

@book{kittel2004introduction,
  title={Introduction to Solid State Physics},
  author={Kittel, C.},
  isbn={9780471415268},
  lccn={2004042250},
  url={https://books.google.co.in/books?id=kym4QgAACAAJ},
  year={2004},
  publisher={Wiley}
}

@inproceedings{Grimvall1981TheEI,
  title={The electron-phonon interaction in metals},
  author={G{\"o}ran Grimvall},
  year={1981},
  url={https://api.semanticscholar.org/CorpusID:92794220}
}

@book{ashcroft1976solid,
  title={Solid State Physics},
  author={Ashcroft, N.W. and Mermin, N.D.},
  isbn={9780030839931},
  lccn={lc74009772},
  series={HRW international editions},
  url={https://books.google.co.in/books?id=oXIfAQAAMAAJ},
  year={1976},
  publisher={Holt, Rinehart and Winston}
}

@article{KT1979,
  title = {Zero-and low-field spin relaxation studied by positive muons},
  author = {Hayano, R. S. and Uemura, Y. J. and Imazato, J. and Nishida, N. and Yamazaki, T. and Kubo, R.},
  journal = {Phys. Rev. B},
  volume = {20},
  issue = {3},
  pages = {850--859},
  numpages = {0},
  year = {1979},
  month = {Aug},
  publisher = {American Physical Society},
  doi = {10.1103/PhysRevB.20.850},
  url = {https://link.aps.org/doi/10.1103/PhysRevB.20.850}
}

@article{BrandtIGL,
  title = {Precision Ginzburg-Landau Solution of Ideal Vortex Lattices for Any Induction and Symmetry},
  author = {Brandt, Ernst Helmut},
  journal = {Phys. Rev. Lett.},
  volume = {78},
  issue = {11},
  pages = {2208--2211},
  numpages = {0},
  year = {1997},
  month = {Mar},
  publisher = {American Physical Society},
  doi = {10.1103/PhysRevLett.78.2208},
  url = {https://link.aps.org/doi/10.1103/PhysRevLett.78.2208}
}

@article{SonierV,
  title = {Muon spin rotation measurements of the vortex state in vanadium: A comparative analysis using iterative and analytical solutions of the Ginzburg-Landau equations},
  author = {Laulajainen, M. and Callaghan, F. D. and Kaiser, C. V. and Sonier, J. E.},
  journal = {Phys. Rev. B},
  volume = {74},
  issue = {5},
  pages = {054511},
  numpages = {9},
  year = {2006},
  month = {Aug},
  publisher = {American Physical Society},
  doi = {10.1103/PhysRevB.74.054511},
  url = {https://link.aps.org/doi/10.1103/PhysRevB.74.054511}
}

@article{lian2018topological,
  title={Topological quantum computation based on chiral Majorana fermions},
  author={Lian, Biao and Sun, Xiao-Qi and Vaezi, Abolhassan and Qi, Xiao-Liang and Zhang, Shou-Cheng},
  journal={Proceedings of the National Academy of Sciences},
  volume={115},
  number={43},
  pages={10938--10942},
  year={2018},
  publisher={National Academy of Sciences}
}

@article{babafemi2025,
    author = {Agboola, Babafemi S. and Reyes-Gonzalez, Joaquin E. and Sharma, Sudarshan and Gautreau, Jonah and Luke, Graeme M. and Lagos, Maureen J.},
    title = {Anisotropic behavior of plasmons in kagome metal YCr6Ge6},
    journal = {APL Materials},
    volume = {13},
    number = {2},
    pages = {021115},
    year = {2025},
    month = {02},
    issn = {2166-532X},
    doi = {10.1063/5.0246342},
    url = {https://doi.org/10.1063/5.0246342},
    eprint = {},
}

@article{TransportPRX2022,
  title = {Formation of an Electron-Phonon Bifluid in Bulk Antimony},
  author = {Jaoui, Alexandre and Gourgout, Adrien and Seyfarth, Gabriel and Subedi, Alaska and Lorenz, Thomas and Fauqu\'e, Beno\^{\i}t and Behnia, Kamran},
  journal = {Phys. Rev. X},
  volume = {12},
  issue = {3},
  pages = {031023},
  numpages = {9},
  year = {2022},
  month = {Aug},
  publisher = {American Physical Society},
  doi = {10.1103/PhysRevX.12.031023},
  url = {https://link.aps.org/doi/10.1103/PhysRevX.12.031023}
}

@misc{ musrdata,
  author = {https://musr.ca/mud/runSel.html}}
\end{document}